\documentclass[aps,amssymb,amsmath,preprint,showpacs]{revtex4}
\usepackage{graphicx}
\newcommand{\ket}[1]{\left|{#1}\right\rangle}
\newcommand{\bra}[1]{\left\langle{#1}\right|} 
\newcommand{\modu}[1]{\left|{#1}\right|}
\newcommand{\inner}[3]{\left\langle{#1}\right|{#2}\left|{#3}\right\rangle}
\newcommand{\expect}[1]{\left\langle{#1}\right\rangle}
\begin{document}
\preprint{IITM/PH/TH/2004/1}
\preprint{quant-ph/0401181}
\title{Signatures of wave packet revival phenomena in the 
expectation values of observables}
\author{C. Sudheesh, S. Lakshmibala, and V. Balakrishnan}
\email{sudheesh,slbala,vbalki@chaos.iitm.ernet.in}
\affiliation{
Department of Physics, Indian Institute of Technology Madras, 
Chennai 600 036, India}
\begin{abstract}
Wave packet revivals and fractional revivals are striking 
quantum interference phenomena that can occur 
under suitable conditions in a system with a nonlinear spectrum.
In the framework of a specific model (the
propagation of an initially coherent wave packet in a Kerr-like
medium), it is shown that distinctive signatures of these
revivals and fractional revivals are displayed by the time evolution
of the expectation values of physical observables and their powers,
i.e., by experimentally measurable quantities. Moreover, different
fractional revivals can be selectively identified by examining
appropriate higher moments.   
\end{abstract}
\maketitle

\section{Introduction}
It is well known that the dynamics of a quantum wave packet whose evolution is
 governed by a nonlinear Hamiltonian can display a variety of non-classical 
effects\cite{wall},\cite{drum}, arising from the 
interference between the stationary states 
comprising the wave packet. These effects manifest 
themselves through several interesting observable 
phenomena such as revivals, fractional revivals and super-revivals of the 
wave packet at specific instants of time\cite{park}-\cite{bann}.
Such revivals have been identified in several physical 
systems. The phenomenon has been 
observed experimentally and
investigated theoretically in a wide class of suitably prepared  
states subject to specific nonlinear potentials, including 
Rydberg wave packets\cite{park}-\cite{meac} and molecular vibrational 
states\cite{vrak},\cite{dool}. For a recent comprehensive review of
 revival phenomena, see \cite{robi}.

Detailed calculations have been made on revival times within the 
framework of the Jaynes-Cummings model of a two-level atom interacting 
with a single mode of electromagnetic radiation\cite{eber},\cite{aver2}.
Full and fractional revivals of angular momentum coherent states\cite{rozm}, 
$SU(1,1)$
coherent states\cite{bann}, states evolving in specific one-dimensional 
potentials like the infinite square well\cite{aron}, and general multi-level 
quantum systems\cite{leic} have also been studied.

The non-classical effects that arise in the dynamics of specific 
wave packets evolving in a Kerr medium\cite{yurk}, collapse and revival 
phenomena displayed by a single-mode field propagating in a 
nonlinear medium\cite{agar}, and the temporal behavior of a macroscopic 
superposition of coherent states governed by a nonlinear 
Hamiltonian\cite{tara} have been reported upon. An
interesting analogy between revivals of wave
packets and Poincar\'e recurrences in classical maps has also been
pointed out, and the interplay between revivals,
recurrences and geometric phases has been elucidated
\cite{sesh1}, \cite{sesh2}.

Much of the attention given to the revival phenomenon has been with a
view to understanding 
the precise nature of the 
transition from quantum to classical dynamics, and the deviations 
from classical predictions that are displayed during the long-time 
evolution of quantum states. Wave packets prepared so as to satisfy the 
minimum uncertainty condition at the initial instant of time are clearly  
appropriate candidates for this purpose. A Rydberg state (an electron 
excited to a high energy level in the bound-state spectrum 
by the application of a suitable laser pulse), for instance, is well 
described by a wave packet which is sharply peaked about a large 
quantum number $n_{0}$, and the spectrum is approximately linear 
in this energy region. Although such a wave packet generally 
spreads with time, it can display, under special 
circumstances, complete or full revivals as well as 
fractional revivals at specific 
instants of time during its evolution.

What has emerged from both experimental and 
theoretical investigations of this phenomenon is the following:
the broad features of wave packet dynamics are quite generic, 
regardless 
of the details of the physical system concerned, the initial 
wave packet 
considered, and the specific nonlinear Hamiltonian 
governing its time evolution. Given an initial state 
$\ket{\psi (0)}$, if the overlap function $C(t)
=\left|\langle \psi(0)|\psi(t)\rangle\right|^2$ returns to its initial
value of unity at specific instants of time, we have a
revival. Revival phenomena are essentially controlled by the
parameters occurring in the first- and second-order terms in the Taylor
expansion of the energy spectrum 
$E_{n}$ about the energy $E_{n_{0}}$ corresponding to the peak of 
the wave packet, namely, 
\begin{equation}
E_{n}-E_{n_{0}}=\sum_{m=1}^{\infty}
\frac{1}{m!} \left[\frac{d^{m}E_{n}}{dn^{m}}\right]_{n=n_{0}}
(n-n_{0})^{m}.
\label{taylorexpn}
\end{equation}
The quadratic term in Eq. (\ref{taylorexpn}) 
can also lead to fractional revivals that occur at 
specific instants between two successive revivals 
of the wave packet. At these times the initial wave packet evolves to a state 
that can be described as a collection of spatially distributed 
sub-packets, each of which closely resembles the initial wave packet. 
Since each of the basis states comprising the wave packet acquires a 
different phase during its temporal evolution, in general, fractional 
revivals can occur only if the numerical 
values of the parameters that appear in the Hamiltonian satisfy
certain conditions, namely, if certain ratios of these 
parameters are sufficiently close to 
rational numbers. Anharmonic corrections due to the 
cubic and higher order contributions in Eq. (\ref{taylorexpn}),
leading to so-called super-revivals,   
have also been 
investigated, in particular in the context of Rydberg atoms
\cite{bluh2},\cite{leic},\cite{knos},\cite{brau1}. 
For wave packets which are 
peaked sufficiently sharply about $E_{n_{0}}$, these higher order
effects that occur on comparatively longer time scales are negligible.

While it is evident that revivals and fractional revivals 
are genuinely quantum 
mechanical in origin, it is of interest to examine the signatures of
these phenomena in experimentally measurable quantities, 
i.e., upon the expectation
values of appropriate physical
observables\cite{brau2}. More specifically, we are concerned 
with the possibility of finding   
distinctive signatures of different fractional revivals occurring
between successive revivals   
that would enable one to distinguish between them unambiguously. 
In this paper, we use the example of a specific nonlinear Hamiltonian
that is both physically relevant and 
for which analytic calculations can be performed, 
to show that this is indeed feasible. This implies that revival phenomena
can be investigated directly, in terms of these 
experimentally accessible expectation values. 

The plan of the paper is as follows: In the next section we briefly 
outline the salient aspects of revivals and fractional 
revivals relevant to our purposes, including  
the relationship between the time of occurrence of a specific fractional 
revival and the number of spatially distributed sub-packets 
representing the state at that instant of time. 
In Section III we obtain expressions 
for the mean and higher moments of the position and momentum operators 
and track their behavior as the wave packet evolves in time. This helps us 
identify the distinguishing features displayed by these 
quantities at the instants when fractional revivals occur. 
We conclude with a discussion of our results in Section IV.

\section{Revivals and fractional revivals}

Consider a nonlinear time-independent Hamiltonian 
$H$ with a non-degenerate spectrum $\{E_n\}$ and eigenstates 
$\{\ket{\phi_n}\}$. The initial state 
$\ket{\psi(0)}$             
in which the system is prepared is a superposition of these  
eigenstates, sharply peaked about some $n_{0}$. 
For ease of notation we shift $n$ by $n_0$ in Eq. (\ref{taylorexpn}), 
and retain terms up to second order in $r = n-n_{0}$, since only these terms 
contribute to revivals and fractional revivals. 
The unitary time evolution operator 
$U(t)=\exp\,(-iHt/\hbar)$ can be written 
in the natural basis in standard notation\cite{robi} as 
\begin{equation}
U(t)=\sum_{r}\exp\,\left\{-2\pi i t \left(\frac{r}{T_{\rm cl}}
 + \frac{r^2}{T_{\rm rev}}\right)\right\}\,
\ket{\phi_r}\bra{\phi_r},
\label{timedevelop}
\end{equation}
apart from an overall phase
factor. Here $T_{\rm cl} = h/|E'(n_0)|$ and $T_{\rm rev} =
2h/|E''(n_0)|$. 
The quadratic term in the exponent  
is responsible for the spreading and distortion 
of the initial wave packet. However, if there exist 
specific instants of time such that, for every $r$ in the summation,  
the exponent is equal to an integer multiple of  $2\pi i$,
then $U(t)$ reduces to the identity operator apart from an 
overall phase factor, and a revival occurs. Further, as mentioned in
Sec. I, fractional revivals can occur at certain instants of time in
between successive revivals. In general, exact revivals require
special values of the ratio $T_{\rm cl}/T_{\rm rev}$, but 
generically 
$C(t)$ can take on values arbitrarily close to unity 
at certain instants of time\cite{sesh1}. In some sense, 
such revivals may be regarded as analogs of Poincar\'{e} 
recurrences of a classical dynamical system 
in a coarse-grained phase space.

For definiteness, we shall consider in the rest of this paper the
example of wave packet evolution in a Kerr-like medium. The
relevant physics of this situation is captured by the model
Hamiltonian\cite{agar},\cite{kita}   
\begin{equation}
H = \hbar\chi \,a^{\dagger 2}\,a^2=\hbar\chi \,N(N-1)
\label{hamiltonian}
\end{equation}
in the usual
notation, with $N = a^{\dagger}\,a,\, \chi = \,$ positive constant.
Essentially the same spectrum models\cite{grei} the interaction
energy of the atoms in a Bose-Einstein condensate. 
The initial state $\ket{\psi(0)}$ is taken to be 
the oscillator coherent state $\ket{\alpha}$ (where $\alpha \in
\mathbb{C}$),  
as it is a minimum position-momentum uncertainty state, and hence is
``classical'' in this sense.  
The corresponding position-space and  momentum-space wave functions
are Gaussian wave packets. Owing to the nonlinear nature of $H$,
these wave packets not only 
spread out as $t$ increases, but also {\em change shape}, 
and the state loses its 
coherence property. We have  
$U(t)=\sum_{0}^{\infty} \exp\,[-i\chi n(n-1)t]\ket{n}\bra{n}$, 
where $\ket{n}$ denotes the usual number
operator eigenstate. Since $n(n-1)$ is always an even integer, it is
evident that $T_{\rm cl} = T_{\rm rev} = \pi/\chi$ in this case. 

In between $t =0$ and $t = T_{\rm rev}$,  
fractional revivals occur at times $t= \pi l/m\chi$, where 
$m = 2,3,4,\ldots$,  and $l = 1, 2,\ldots, (m-1)$ for a given value of
$m$. This can be traced to the 
interesting periodicity property exhibited by $U$ at these 
instants of time, 
consequent to which it can be Fourier expanded in the basis set 
$\{\exp\,(-2\pi il/m),\,0\leq l \leq (m-1)\}$ for each $m$. 
Writing 
\begin{equation}
\ket{\psi (t)}= 
e^{-\modu{\alpha}^{2}/2}\sum_{n=0}^{\infty}\frac{e^{-i\chi 
n(n-1)t} \,\alpha^n}{\sqrt{n!}}\ket{n}
\label{psit}
\end{equation}
we see that,  at these instants of time, 
the exponential in the summand 
can be absorbed 
in the complex number $\alpha$. Thus $\ket{\psi(\pi/m\chi)} 
\equiv \ket{\psi_m}$ 
becomes a {\em finite} superposition of ``rotated'' coherent states with 
definite amplitudes, according to 
\begin{equation}
\ket{\psi_m}=\left\{\begin{array}{ll}
\sum_{l=0}^{m-1} f_{l}\ket{\alpha \,e^{-2\pi il/m}},\quad& 
m \,{\rm odd} \\[10pt]

\sum_{l=0}^{m-1} g_{l}\ket{\alpha \,e^{i\pi/m}\,e^{-2\pi il/m}},\quad& 
m \,{\rm even,} \end{array} \right.
\end{equation}
where $f_l$ and $g_l$ are the Fourier coefficients. For instance, 
the state at time 
$\pi/(2\chi)$ (corresponding to $m = 2$) 
is a superposition of the two 
coherent states $\ket{i\alpha}$ and $\ket{-i\alpha}$. In general, 
$\ket{\psi_m}$ is a superposition of $m$ coherent states. The
corresponding wave packet in position space is 
a superposition of $m$ spatially distributed Gaussian wave packets, 
and a fractional revival is said to have 
occurred. The periodicity property of $U$ further implies 
that the wave function at times $t=\pi l/m \chi,\, 1\leq l \leq m-1,$ 
is also a superposition of $m$ wave packets\cite{aver1}.
It is therefore clear that experimental observations  
of fractional revivals that depend on probing the wave packet 
itself are quite involved and somewhat indirect, 
and careful interpretation of the results is required before firm
conclusions can be drawn. It is this aspect that leads us to suggest
that an examination of the moments of observables may provide a
less intricate and more definitive means of identifying and sorting
out revival phenomena.   

\section{Moments of observables}
We now show how fractional revivals are 
mirrored, in distinctive ways, in the 
expectation values of the physical observables
pertaining to the system. As the system enjoys 
revivals with a period $T_{\rm rev}$, all such expectation values are 
periodic functions of $t$ with this fundamental period. 

The relevant observables are the position $x = (a +
a^{\dagger})/\sqrt{2}$ and $p =(a -
a^{\dagger})/i\sqrt{2}$.  
Clearly, their expectation values alone do not suffice to reproduce the full
information contained in the wave function itself. In principle, 
an infinite set of
moments, namely, the expectation values of {\em all} powers of $x$ and $p$ and
their combinations, is required for this purpose.  In this sense, the
quantum system is equivalent to an infinite-dimensional classical
system in which the role of the dynamical variables is played by the
set of expectation values. However, we emphasize that even the first few
moments can be seen to yield considerable
information on the behavior of the system. 

Recalling that $a\ket{\alpha} = \alpha \ket{\alpha}$, it is
convenient to define the $c$-number
function 
\begin{equation}
\alpha(t)=\inner{\psi(t)}{a}{\psi(t)}=
\inner{\alpha}{e^{iHt/\hbar}\,a\,e^{-iHt/\hbar}}{\alpha},
\label{alphat1}
\end{equation}
so that $\alpha (0) \equiv \alpha$. 
In the special case in which $H = H(N)$, this simplifies to the
explicit expression
\begin{equation}
\alpha(t)=\alpha \,e^{-\modu{\alpha}^2}
\sum_{n=0}^{\infty} \frac{\modu{\alpha}^{2r}}{r!}\,
e^{-i[H(r+1)-H(r)] t/\hbar}\,.
\label{alphat2}
\end{equation}
In general, this is a fairly complicated function of $t$. For the case at 
hand, however, with $H=\hbar\chi N(N-1)$, 
some simplification occurs. We find
\begin{equation}
\alpha(t)=\alpha \,e^{ -\modu{\alpha}^2(1-\cos 2\chi t)}\, 
\left[ \cos \,\big(\modu{\alpha}^2\,\sin \,(2\chi t)\big)
-i\sin\,\big(\modu{\alpha}^2\,\sin \,(2\chi t)\big)\right].
\label{alphat3}
\end{equation}
Thus $\alpha(t)$ is a periodic function of time with period 
$\pi/\chi$, as expected.

It is convenient to introduce the notation 
$\alpha =\alpha_1+i\alpha_2 \equiv (x_0 + i p_0)/\sqrt{2}$ and 
$\nu=\modu{\alpha}^2 = \frac{1}{2}(x_{0}^{2}+p_{0}^{2})$.
As is well known, $x_0$ and $p_0$ represent the locations of the 
centers of the initial Gaussian wave packets in position and momentum
space, respectively.  
Then the expectation values of $x$ and $p$ can be written as explicit
functions of $t$ in the form
\begin{equation}
\expect{x(t)}=
e^{-\nu\,(1-\cos 2\chi t)}\,
\big[ x_0 \,\cos \,(\nu \sin 2\chi t)+p_0\, 
\sin \,(\nu \sin 2\chi t)\big],
\label{xt}
\end{equation}
\begin{equation}
\expect{p(t)}=
e^{-\nu(1-\cos 2\chi t)}\,
\big[ -x_0 \,\sin\, (\nu \sin 2\chi t)+p_0\, 
\cos \,(\nu \sin 2\chi t)\big].
\label{pt}
\end{equation}
Similarly, the second moments of $x$ and $p$ are found 
easily from the expression
$\expect{a^2(t)}=\alpha^{2}\, \exp\,\big(-2i\chi t -\nu
+ \nu\, e^{-4i\chi t}\big)$
and its complex conjugate. We get
\begin{eqnarray}
2 \expect{x^2(t)}&=& 1 
+ x_{0}^{2} + p_{0}^{2} + e^{-\nu\,(1-\cos \,4\chi t)}
\,\big[ (x_{0}^{2} - p_{0}^{2})\,\cos\,  
(2\chi t + \nu \sin \,4\chi t)\nonumber\\ 
&+& 2 x_{0} p_{0} \,\sin\, (2\chi t + \nu\sin \,4\chi t)\big],
\label{xsqrdt}
\end{eqnarray}
\begin{eqnarray}
2 \expect{p^2(t)}&=& 1 
+ x_{0}^{2} + p_{0}^{2} - e^{-\nu\,(1-\cos \,4\chi t)}
\,\big[ (x_{0}^{2} - p_{0}^{2})\,\cos\,  
(2\chi t + \nu \sin \,4\chi t)\nonumber\\ 
&+& 2 x_{0} p_{0} \,\sin\, (2\chi t + \nu\sin \,4\chi t)\big].
\label{psqrdt}
\end{eqnarray}
For reasons already mentioned, the higher moments also carry much
information of direct interest. 
The third moments can be written compactly in the form 
\begin{eqnarray}
4\expect{x^3(t)}&=&  e^{-\nu 
(1-\cos 6\chi t)}\big[(x_0^3-3x_0 p_0^2)
\,\cos\, (6\chi t+\nu\sin 6\chi t) +(3x_0^2 p_0- p_0^3)\nonumber\\ 
&\times&\sin \,(6\chi t+\nu\sin 6\chi t)\big]
+6\nu \,\big[\expect{x(t)}(1 + 
\cos\, 2\chi t) + \expect{p(t)} 
\,\sin\,2\chi t)\big],
\label{xcubedt}
\end{eqnarray}
\begin{eqnarray}
4\expect{p^3(t)}&=&  e^{-\nu 
(1-\cos 6\chi t)}\big[(x_0^3-3x_0 p_0^2)\,\sin\, (6\chi 
t+\nu\sin 6\chi t) -(3x_0^2 p_0- p_0^3)\nonumber\\
&\times&\cos \,(6\chi t+\nu\sin 6\chi t)\big]
+6\nu\,\big[\expect{p(t)} (1+\cos 2\chi t)-
\expect{x(t)}\,\sin \,2\chi t\big].
\label{pcubedt}
\end{eqnarray}
With these expressions and the explicit expressions 
for the fourth moments (which we do not write them down here), 
the variances
of $x$ and $p$ as functions of $t$, as also  
the skewness and kurtosis in each case,  
can be obtained.
The uncertainty product (or the product of the standard deviations)
$\Delta x \,\Delta p$, which initially has the minimum value
$\frac{1}{2}$, is of special interest. We do not write down the
lengthy expressions for these quantities here, but we shall comment upon
their time variation in the sequel.
Finally, we note that expressions for the higher 
moments can be deduced readily from the general result
\begin{equation}
\expect{a^{\dagger k}\,a^{k+l}} = \alpha^{l}\,\nu^{k}
e^{-\nu\,(1-\cos \,2 l \chi t)}\,
\exp \left[
-i \chi \big(l(l-1) + 2kl\big)\,t - i\nu \,\sin\,2l \chi t\right],
\label{nthmoment}
\end{equation}
where $k$ and $l$ are non-negative integers.  

We now turn to a discussion of our results. This is most conveniently
done with the help of 
plots drawn for typical values of the parameters
concerned.

\section{Results and Discussion}

We first compare 
the explicit solutions found for the expectation
values of $x$ and $p$ in Eqs. (\ref{xt}) and (\ref{pt}) with the
solutions that would have been obtained for $x(t)$ and $p(t)$
{\it had} the system been a {\it classical} one, governed by the classical
counterpart of the normal-ordered Hamiltonian  
$H =\hbar\chi \,a^{\dagger 2}\,a^2$, namely, 
$H_{\rm cl} = \frac{1}{4} (x^2 + p^2)^2$. 
Although the equations of motion corresponding to $H_{\rm cl}$ are
nonlinear, it is evident that $x^2 + p^2$ is a constant of the motion,
so that the phase trajectories are circles. However, the 
frequency of motion is dependent on the initial conditions (i.e., the
amplitude of the motion), being
equal to $\nu = \frac{1}{2}(x_{0}^2 + p_{0}^2)$. This is, of course,
a well-known feature of nonlinear oscillators. But we note that
the actual solutions
for $\expect{x(t)}$ and $\expect{p(t)}$ in Eqs. (\ref{xt}) and
(\ref{pt}) are more complicated than the classical ones for $x(t)$ and
$p(t)$ under $H_{\rm cl}$. This is a consequence of the 
quantum mechanical nature of the system, over and above the
nonlinearity of $H$. 
However, the expressions for $\expect{x(t)}$ and $\expect{p(t)}$ can
be given the following 
interesting interpretation in classical terms. 
Define the (non-canonical) pair of classical dynamical variables 
\begin{equation}
X = x\,e^{\nu\,(1-\cos 2\chi t)}\,,\quad
P = p\,e^{\nu\,(1-\cos 2\chi t)},
\label{XPdefn}
\end{equation}
and the {\em reparametrized time} $\tau = \sin\, (2\chi t)$. 
The initial values $X_{0}$ and $P_{0}$ of these variables 
remain equal to 
$x_{0}$ and $p_{0}$, respectively. Then Eqs. (\ref{xt}) and (\ref{pt})
can be re-written as 
\begin{equation}
X = X_{0}\,\cos\,\nu\tau + P_{0}\,\sin\,\nu\tau\,,\quad
P = -X_{0}\,\sin\,\nu\tau + P_{0}\,\cos\,\nu\tau.
\label{XPtau}
\end{equation} 
But these are the solutions to the system of equations
\begin{equation}
\frac{dX}{d\tau} = \nu P\,,\,\, \frac{dP}{d\tau} = -\nu X\,,\,\, 
\label{}
\end{equation}
describing a nonlinear oscillator of frequency $\nu = 
\frac{1}{2}(X_{0}^2 + P_{0}^2)$, in terms of
the transformed variables $(X,P)$ and the reparametrized time $\tau$. 
At the level of the {\em first} moments, therefore, the system is
effectively a nonlinear oscillator after a suitable transformation
of the relevant variables.

Turning to the details of the time dependence of the various moments of
$x$ and $p$, there are two striking features 
that underlie the essential point we wish to make in this paper. 
First, the higher the order of the moment (or cumulant), the 
more rapid is its variation, since the leading frequency 
in the $m^{\rm th}$ moment is $2m\chi$.
Second, the time dependence is strongly controlled by the 
factor $\exp\,[-\nu\,(1-\cos \,2 m \chi t)], \, m = 1,2,\ldots,  $ that 
modulates the oscillatory terms. While this permits substantial time
variation for sufficiently small values of $\nu$, it acts as a strong
damping factor for large values of $\nu$, {\it except when
$\cos\,(2m\chi t)$ is near unity}. As one might expect, this happens precisely 
at revivals (when $t = n\pi/\chi$, an integer multiple of $T_{\rm
rev}$), as expected. But it also happens, in the $m^{\rm th}$ moment
alone, at the fractional revival
times $t = (n + l/m)T_{\rm rev}$. Thus, by setting $\nu$ at a suitably
large value, we can ensure that the moments are essentially static,
bursting into rapid variation at specific instants of time before
reverting to quiescence. 

These points are illustrated by the figures that follow.
Owing to an obvious symmetry of the Hamiltonian, the
moments of $x$ and $p$ behave in an essentially similar manner,
especially if we start with the symmetric initial condition 
$x_0 = p_0$. Without loss of generality, we restrict ourselves 
to this case in what follows. We have set $\chi = 5$ in the numerical 
results to be presented, but this is irrelevant as all the plots
correspond to $t$ measured in units of $\pi/\chi$. 
We find that for very small values ($ \ll 1$) of $x_{0}$ and $p_{0}$
(i.e., of $\nu$), the
nonlinearity of $H$ does not play a significant role, and the behavior
of the system is much like that of a simple oscillator. Interesting
behavior occurs for larger values of $\nu$. We therefore present
results for three typical values of the parameters
representing the initial conditions, namely:
(a) $x_0 = p_0 = 1 \Rightarrow \nu = 1$; 
(b) $x_0 = p_0 = \sqrt{10} \Rightarrow \nu = 10$; 
(c) $x_0 = p_0 = 10 \Rightarrow \nu = 100$. 
These correspond, respectively, to small, intermediate, 
and large values of $\nu$.
In all the ``phase plots'', the point representing the state at $t =
0$ is labeled A. 

\begin{figure}
\includegraphics[width=7in]{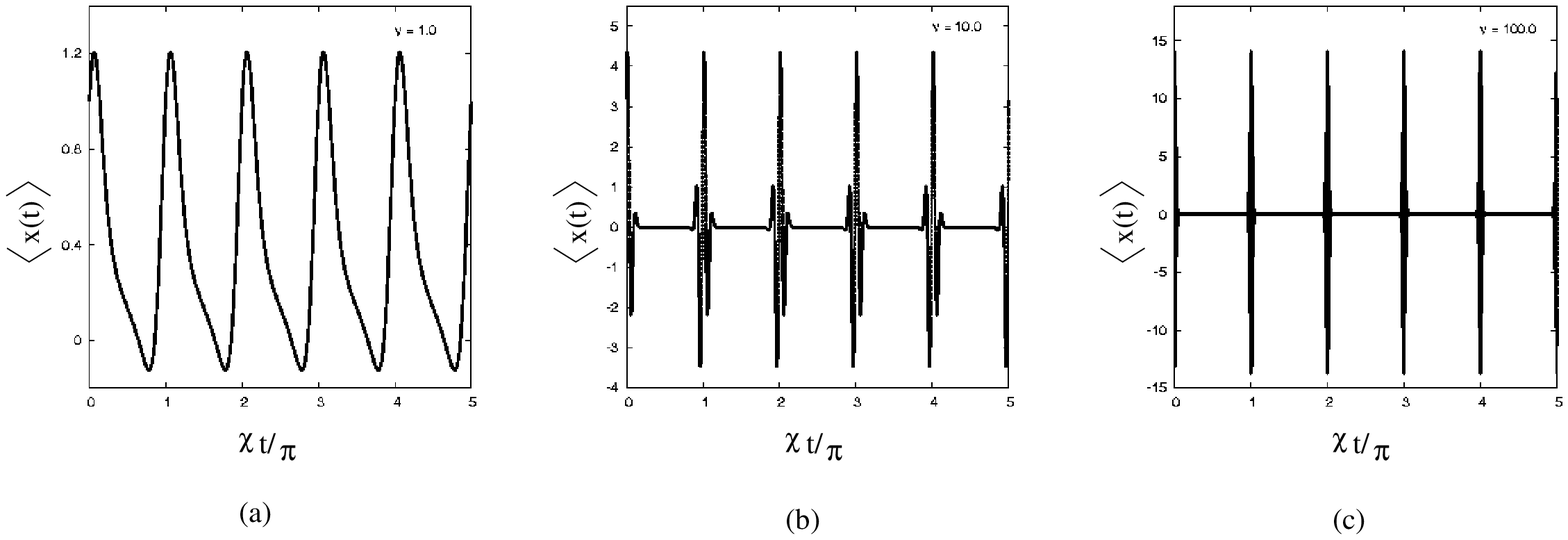}
\caption{$\expect{x}$ as a function of time}
\label{fig1}
\end{figure}
Figures 1(a)-(c) show the variation of $\expect{x(t)}$ 
as a function of $t$ for small, medium and large values of $\nu$.
(As already mentioned, $\expect{p(t)}$ displays similar behavior.)
For sufficiently large values of $\nu$, it is evident
that, except for times close to integer multiples of 
$T_{\rm rev} = \pi/\chi$, 
$\expect{x(t)}$ and $\expect{p(t)}$ essentially remain static at the
value zero. 
\begin{figure}
\includegraphics[width=7in]{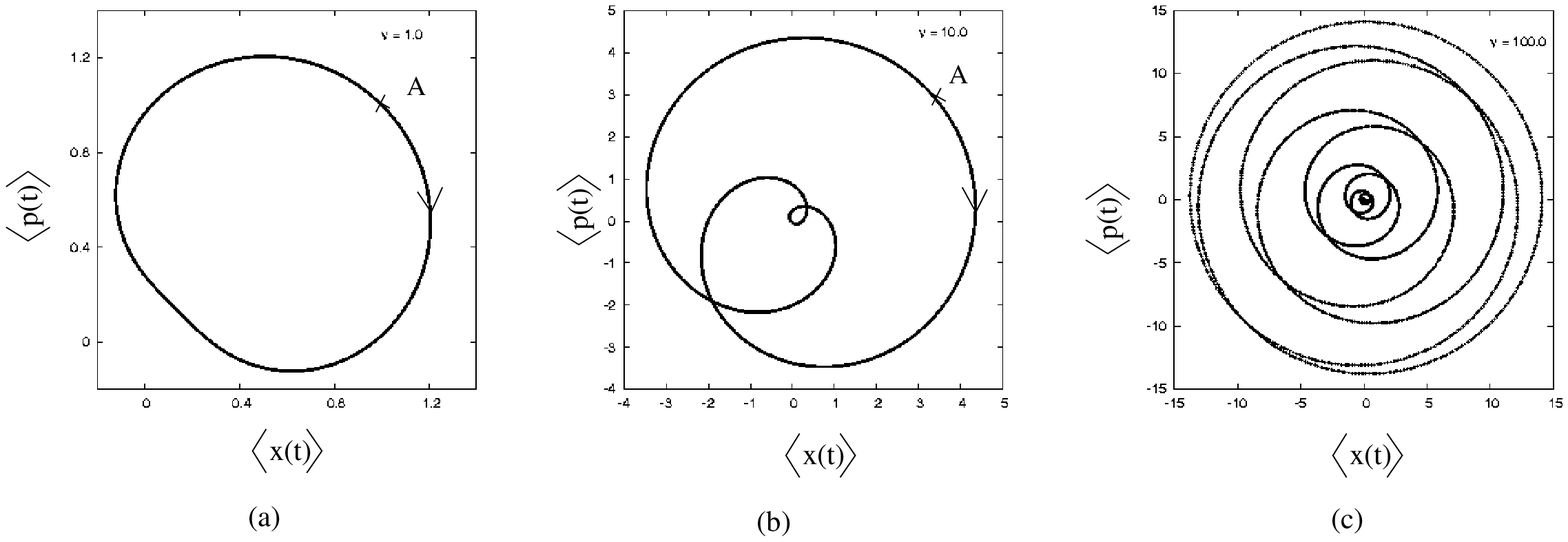}
\caption{``Phase plot'' of $\expect{p}$ versus $\expect{x}$}
\label{fig2}
\end{figure}
Figures 2(a)-(c) depict the corresponding 
``phase plots'' in the $(\expect{x}, \expect{p})$ plane. 
In Fig. 2(c), the representative point remains at the origin  
most of the time, except at times close to successive revivals,
when it rapidly traverses the rest of the curve before returning to
the origin. 

While sudden changes from nearly static values of $\expect{x(t)}$ 
and $\expect{p(t)}$ are thus signatures of revivals, the 
occurrence of fractional revivals is not captured in these mean 
values. The fractional revival 
occurring mid-way between successive revivals (e.g., at $t=\pi/2\chi$ 
in the interval between $t = 0$ and $t = T_{\rm rev}$),  
when the initial wave packet reconstitutes itself into two separate 
wave packets of a similar kind, 
leaves its signature upon the second moments.
Figures 3(a)-(c) show the variation with time of the uncertainty product 
$\Delta x\, \Delta p$. 
In each case, this product returns to its
initial, minimum value $\frac{1}{2}$ at every revival, rising to higher
values in between revivals. 
\begin{figure}
\includegraphics[width=7in]{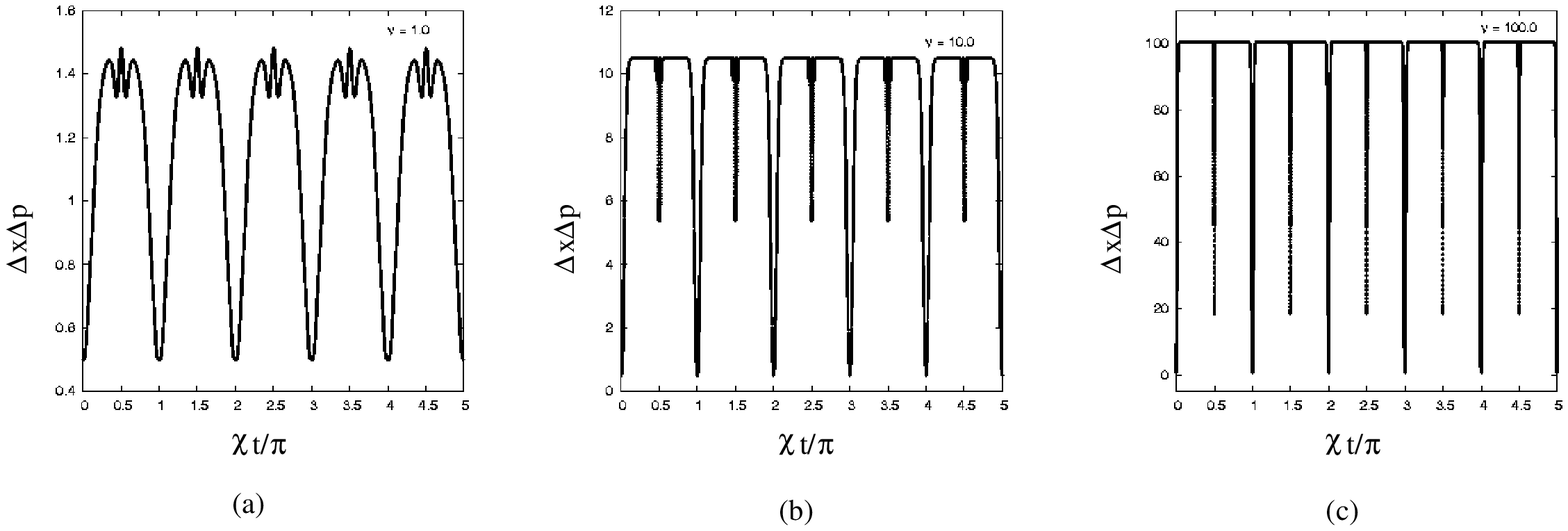}
\caption{Variation of the uncertainty product with time}
\label{fig3}
\end{figure}
Once again, for sufficiently large values
of $\nu$, the product remains essentially static at the approximate
value $(\frac{1}{2} + \nu)$ for most of the time, but undergoes   
extremely rapid variation near revivals, and {\em also} near the fractional 
revivals occurring mid-way between revivals. 
During the latter,  
the uncertainty product drops to smaller values, but does not
reach the minimum value $\frac{1}{2}$. 

There is a very striking 
difference in the behavior of the standard deviations 
near revivals as opposed
to their behavior near the foregoing fractional revivals. 
This is brought out in Figs. 4(a)-(c),
which is a ``phase plot'' of $\Delta p$ versus $\Delta x$.  
\begin{figure}
\includegraphics[width=7in]{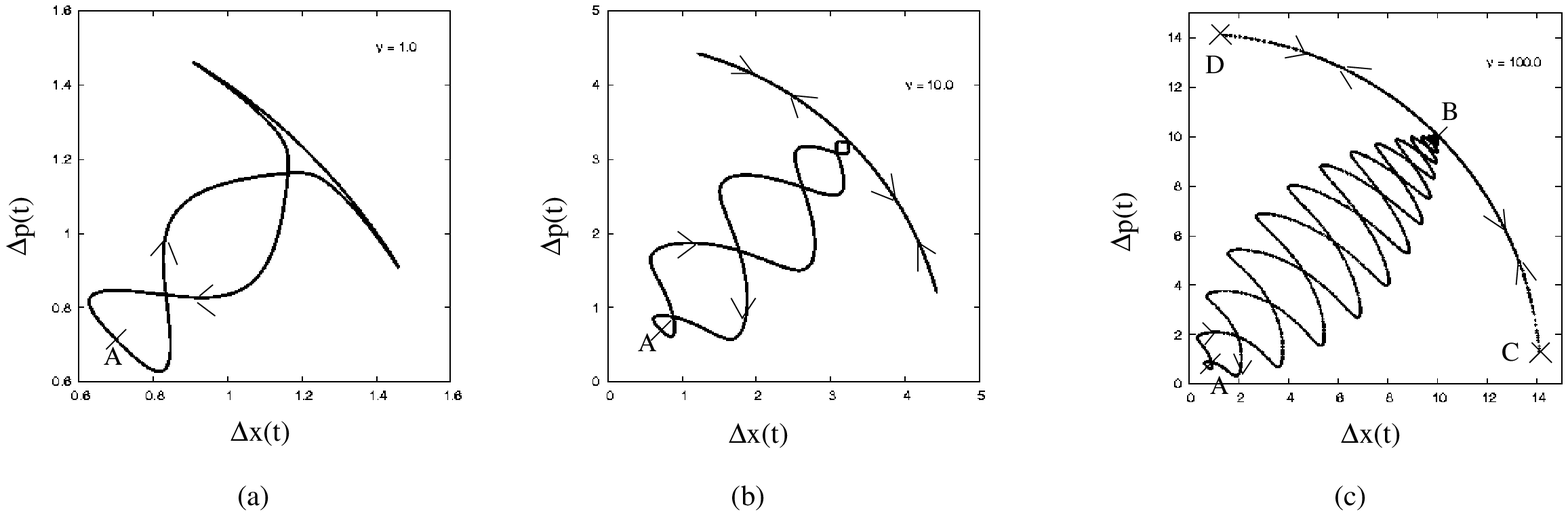}
\caption{``Phase plot'' of $\Delta p$ versus $\Delta x$}
\label{fig4}
\end{figure}
For very small $\nu$, as in Fig. 4(a), $\Delta x$ and $\Delta p$ vary quite
gently around a simple closed curve. When $\nu$ is somewhat larger, as
in \ref{fig4}(a) which corresponds to $\nu = 1$, the plot begins to
show interesting structure. 
For much larger values of $\nu$ as in \ref{fig4}(c), 
the initial point A quickly moves out
on the zig-zag path about the radial $\Delta p = \Delta x$ line
to the steady value represented by the point
B, and returns to A at every revival along the
complementary zig-zag path. 
Close to the fractional revival at $t = (n+\frac{1}{2})T_{\rm rev}$,
however, the representative point moves back and forth along the 
azimuthal path
BCDB rather than the zig-zag path: clearly, a kind of ``squeezing''
occurs, as one of the variances reaches a small value while the other
becomes large, and vice versa. (Of course the state of the system 
is far from a minimum uncertainty state throughout, except
at the instants $n T_{\rm rev}$.)  

The fractional revivals occurring at 
$t = (n + \frac{1}{3})T_{\rm rev}$ 
and $t = (n + \frac{2}{3})T_{\rm rev}$, when the initial wave packet 
is reconstituted into a superposition of three separate wave packets, 
are detectable in the third
moments of $x$ and $p$. To make this unambiguous, we may consider the
third moments about the mean values --- or, in standard statistical
notation, the square of the skewness, defined as 
\begin{equation} 
\beta_{1}^{(x)}
= \left\langle\big(x-\expect{x}\big)^3\right\rangle^2\Big/
\big(\expect{x^2}-\expect{x}^2\big)^3,
\label{skew}
\end{equation}
and similarly for $\beta_{1}^{(p)}$. Figures 5(a)-(c) show the
variation of $\beta_{1}^{(x)}$ with $t$. 
\begin{figure}
\includegraphics[width=7in]{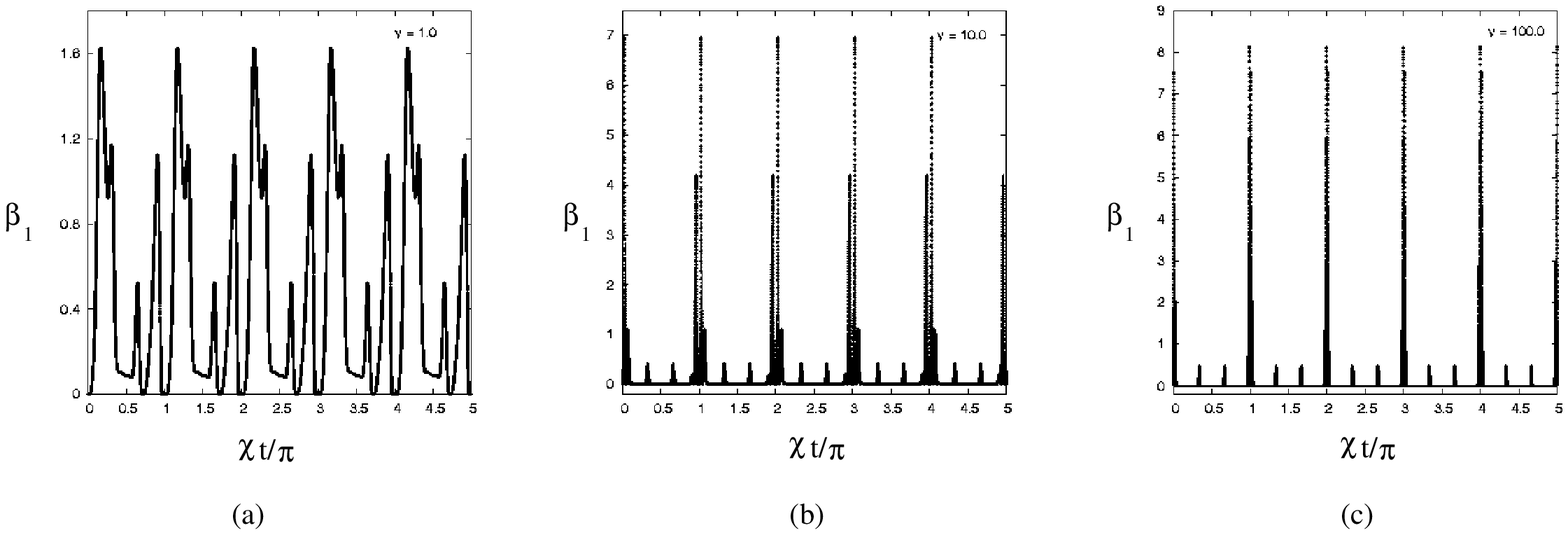}
\caption{Square of the skewness in $x$ as a function of time}
\label{fig5}
\end{figure}
It is evident that, for
sufficiently large values of $\nu$, $\beta_{1}^{(x)}$ remains nearly
zero most of the time, except for bursts of rapid variation close to
revivals and fractional revivals. 
Both $\beta_{1}^{(x)}$ and $\beta_{1}^{(p)}$ actually vanish 
at $t = nT_{\rm rev}$, but they remain non-zero at $t = (n +
l/3)T_{\rm rev}\,, l = 1,2$. More detailed information is obtained
from a ``phase plot'' of $\beta_{1}^{(p)}$ versus $\beta_{1}^{(x)}$,
which we do not give here.

\begin{figure}
\includegraphics[width=7in]{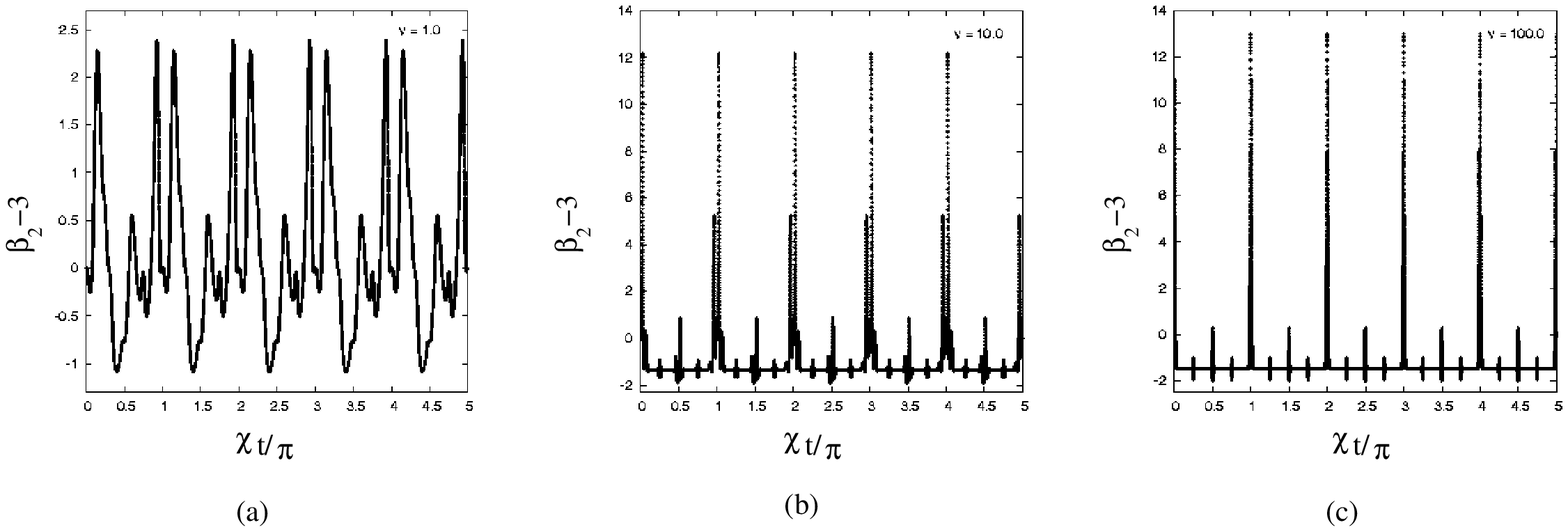}
\caption{The excess of kurtosis of $x$ as a function of time}
\label{fig6}
\end{figure}

Finally, we consider fractional revivals corresponding to 
$m = 4$, when four superposed 
wave packets appear.  
These are
detectable in the behavior of the fourth moments of $x$ and $p$. 
Equivalently, we may use the excess of kurtosis $(\beta_{2} - 3)$,  
where the kurtosis of $x$ is defined as 
\begin{equation}
\beta_{2}^{(x)} 
=\left\langle\big(x-\expect{x}\big)^4\right\rangle\Big/
\big(\expect{x^2}-\expect{x}^2\big)^2,
\label{kurt}
\end{equation}
with a similar definition for $\beta_{2}^{(p)}$. The excess of
kurtosis is the measure of the departure of a distribution from gaussianity.  
Figures \ref{fig6}(a)-(c) depict how $(\beta_{2}^{(x)} -3)$ varies with time.  
For sufficiently large $\nu$, both 
$(\beta_{2}^{(x)} -3)$ and $(\beta_{2}^{(p)}-3)$
remain essentially static near the value 
$-\frac{3}{2}$ for most of the time. They 
vary rapidly near revivals, vanishing 
at $t = n\,T_{\rm rev}$ because the
wave packet is a Gaussian both in position space and in momentum space 
at these instants of time. As is clear from Fig. 4(c), they 
also vary rapidly near the fractional revivals
at $t = (n + l/4)T_{\rm rev}\, ({\rm where} \, l = 1, 2, 3)$, 
oscillating about the
``steady value'' $-\frac{3}{2}$. Once again, a ``phase plot'' of    
$(\beta_{2}^{(p)} -3)$ versus $(\beta_{2}^{(x)}-3)$ (which we do not
give here) helps identify
features that distinguish between the three fractional revivals
concerned.  

We have shown that distinctive, experimentally detectable signatures of 
the different fractional revivals of a suitably prepared 
initial wave packet  
are displayed in the expectation values of physical observables and
their powers. The complicated quantum interference effects that lead
to fractional revivals can thus be captured in the dynamics of these
expectation values, which may be regarded as the dynamical variables  
in a classical phase space.
While this is, in principle, an infinite-dimensional space, what is 
relevant in practice is the temporal behavior of a finite number of 
moments, since fractional revivals corresponding to very 
large values of $m$ are not easy to detect in any case.

We thank Suresh Govindarajan for discussions and
help. This work was supported in part by the
Department of Science and Technology, India, under Project No.  
SP/S2/K-14/2000.

\end{document}